# Make Silence Speak for Itself: a multi-modal learning analytic approach with neurophysiological data

Mingxuan Gao, Jingjing Chen, Yun Long, Xiaomeng Xu, Yu Zhang[*]

**Abstract**

*Background:* Silence is a common phenomenon in classrooms, yet its implicit nature limits a clear understanding of students' underlying learning statuses.

*Aim:* This study proposed a nuanced framework to classify classroom silence based on class events and student status, and examined neurophysiological markers to reveal similarities and differences in silent states across achievement groups.

*Sample:* The study involved 54 middle school students during 34 math lessons, with simultaneous recordings of electroencephalogram (EEG), electrodermal activity (EDA), and heart rate signals, alongside video coding of classroom behaviors.

*Results:* We found that high-achieving students showed no significant difference in mean EDA features between strategic silence (i.e., students choose silence deliberately) and active speaking during open questioning but exhibited higher EEG high-frequency relative power spectral density (RPSD) during strategic silence. In structural silence (i.e., students maintain silence following an external command) during directed questioning, they demonstrated significantly higher heart rates while listening to lectures compared to group activities, indicating heightened engagement. Both high- and medium-achieving students displayed elevated heart rates and EDA tonic components in structural silence during questioning compared to teaching. Furthermore, high-achieving students exhibited lower high-frequency RPSD during structural silence than strategic silence, a pattern not observed in other groups, highlighting group heterogeneity.

*Conclusions:* The findings contribute to validating the complexity of silence, challenge its traditional association with passivity, and offer a novel classification framework along with preliminary empirical evidence to deepen the understanding of silent learning behaviors in classroom contexts.

# 1. Introduction

Students' silence, commonly understood as an absence of verbal communication in

---

[*] Corresponding author. School of Education, Tsinghua University, Beijing, 100084, China.

*Email addresses:* gmx22@mails.tsinghua.edu.cn(Mingxuan Gao), chen-jj15@tsinghua.org.cn (Jingjing Chen), longyun922@mail.tsinghua.edu.cn (Yun Long), xuxm23@mails.tsinghua.edu.cn (Xiaomeng Xu), zhangyu2011@tsinghua.edu.cn (Yu Zhang).

classrooms, has often been labeled as a sign of passive learning (Hein, 1991; Zhouyuan, 2016; Juma et al., 2022). However, this common perspective overlooks essential cognitive activities, such as thinking, imaging, and mind wandering, which fundamentally manifest as silence and play a significant role in learning (Zembylas and Michaelides, 2004; Jaworski, 1992). Due to the implicit nature of these activities, educators and researchers often struggle to differentiate students' learning status during silence, hindering both theoretical development and the selection of effective teaching strategies (Bao, 2020). Thus, re-examining and exploring silence is necessary for advancing both theory and educational practice.

Silence is a complex phenomenon determined by diverse contextual and individual factors. In conventional classrooms, silence may be imposed as an external requirement, symbolizing order and compliance, and indicating the structural nature of the learning environment (Tang et al.,2020; Kim, 2024). Students may wait for reinforcement or further cues from the teacher during silence (Schunk, 1982). In contrast, students also actively choose to remain quiet even in classrooms that encourage verbal interaction (Hu and Fell-Eisenkraft, 2003; Basso, 2013). Additionally, from the perspective of cognitive theory, silence can occur during deep information processing or as a result of cognitive overload (Festinger, 1957; Ling, 2003; Ben-Soussan et al., 2023). In these moments, students might need silence to organize their thoughts and handle the demands of new information. At the same time, individual differences play an important role: high-achieving students might choose silence due to introversion, a preference for listening, or deeper cognitive engagement (Schultz, 2010), while lower-achieving students' silence could stem from disinterest, social anxiety, or low self-confidence (Sedova and Navratilova, 2020). These variations imply that silence cannot be viewed simplistically but rather as a dynamic and multifaceted phenomenon, shaped by a range of contextual and individual influences.

From a constructivist perspective, learning is an active and individual process (Nie and Lau, 2010). Students may engage in diverse implicit activities in silence depending on contextual factors such as motivational drivers and cognitive development (Tang et al., 2020). Reconsidering the traditionally passive stereotype associated with silence reveals its potential as an active learning status. Picard (1952) posited that silence does not begin because language ceases. Also, some researchers proposed that silence does not signify an absence of thought but instead may facilitate cognitive activity (Vago and Zeidan, 2016; Shah, 2019; Kroll, 2004; Reda, 2009; Schultz, 2010), challenging the perception that silence equates to disengagement. For instance, silence can provide students with time for reflection, allowing deeper internalization of knowledge (Kim, 2024). Moreover, teachers may purposefully use silent waiting time as a pedagogical strategy, employing it to enhance students' cognitive processing and engagement (Su et al.,2023). Consequently, silence should not merely be considered a static status of non-participation but rather a context-dependent learning phenomenon that demands careful differentiation to appreciate its educational value (Glenn, 2004; Ha and Li, 2014; Fidyk, 2013; Clarke et al., 2021). The complexity of silence should be considered.

Despite growing recognition of the complexity of silence, current research often focuses on describing its occurrence without examining the cognitive and emotional learning status that may accompany it (Hiebert et al., 2003; Helme and Clarke, 2001). A notable limitation is the simplified description of silence, lacking a theoretical framework. Many studies merely summarize silence-related event in classroom settings, without a comprehensive analysis of students' implicit status during these silent events (Kovalainen and Kumpulainen, 2007). Another limitation in current research on silence lies in the challenge of its representations, as silence is marked by an absence of verbal communication (Hiebert et al., 2003). With cognitive activities occurring internally (Helme and Clarke, 2001), traditional methods, such as classroom observations, self-reports, and interviews, face significant obstacles (Su et al., 2023; Schultz, 2010). As a result, researchers emphasize the need to incorporate more measures to enhance the validity and depth of silence studies (Chan et al., 2020).

Advances in neurophysiological techniques offer promising avenues to examine the "black box" of silence in real classroom (Davidesco et al., 2021; Xu and Zhong, 2018). In addition to the neurophysiological studies on silence during meditation (Ben-Soussan et al., 2020), experiments in the lab (Meshulam et al., 2021, Pan et al., 2020) and in the real classroom (Dikker et al., 2017; Chen et al., 2023; Xu et al., 2024, Feng et al., 2025) provide promising potentials of neurophysiological data in investigating implicit states of learning in education settings. For example, Dikker et al. (2017) found that EEG synchronization among students correlated with classroom engagement and social dynamics. Chen et al. (2023) further found different types of EEG coupling with peers, which may reflect the successful learning states, in different disciplines. Besides central nervous system activity, heart rate and electrodermal activity (EDA), which reflect peripheral nervous system responses, also effectively represent important cognitive and emotional process during learning (Horvers et al., 2021; Zhang et al., 2018; Qu et al., 2020; Zhang et al., 2021; Huber and Bannert, 2023). The above findings indicate that various neurophysiological modalities could play a role in enhancing our understanding of silence in the classroom. Nevertheless, to the best of our knowledge, no previous studies have investigated silence in the classroom from a neurophysiological perspective.

Given these insights, this study aimed to propose a more nuanced framework for classifying silence in classrooms, as detailed in Section 2. Then, we collected videos and neurophysiological data during silence in classroom from 34 math lessons in a middle school (56 students) and explored the multimodal representations of silence to uncover its complexity. At the same time, considering the possible influence of individual differences on silence, we further investigated the heterogeneity across groups with different academic performances.

## 2. Framework of Classifying Silence

This section introduces a comprehensive framework for classifying student silence during classroom learning, which serves as the foundation for systematically coding and analyzing classroom video data in this study. Silence in education is a complex and multifaceted phenomenon. From a sociocultural perspective, it may serve as a communicative act that expresses respect, maintains group harmony, or mitigates negative social exposure (Cortazzi & Jin, 2001; Hofstede et al., 2014; Vygotsky, 1978). From a cognitive perspective, silence can reflect internal processing, rehearsal, or deferred participation (Sweller, 1988; Biggs, 1996). Meanwhile, social interaction theory (Goffman, 2023) views silence as part of the face-management repertoire in classroom performances, and self-determination theory (Deci & Ryan, 2013) considers it a possible sign of motivational disengagement or autonomy.

To reconcile these varied interpretations and enable operationalization in video-based data, this study adopts a two-layered framework. First, we categorize classroom discourse into six typical types of events, each reflecting distinct instructional formats and interaction patterns.

Second, we classify student behaviors—especially silence—based on their underlying motivational or institutional drivers, distinguishing between strategic and structural silence. This distinction not only captures the functional diversity of silence but also facilitates reliable annotation using externally observable cues.

**2.1. Class Event Description**

This classification scheme ensures that each observed behavior is interpreted within its specific instructional context, allowing for a more accurate understanding of when and why silence occurs. The following section builds on this event-based categorization by introducing a typology of student silence grounded in both contextual triggers and internal learner processes.

- Question to Directed Student (QtDS): The teacher addresses a specific student, assigning them a speaking role, while others observe.
- Question to All Students (QtAS): Open questions are posed to the class; students respond voluntarily.
- New Content Teaching: Teacher-led instruction presenting new knowledge; typically unidirectional with no student talk.
- Quiz Elaboration: Teacher explains answers to tests or homework; similarly unidirectional but focused on review.
- Group Activity: Students collaborate in small groups, reciting or discussing with peers.
- Quiz: Students complete written work independently, with minimal or no teacher interaction.

These event categories reflect varying degrees of learner agency and verbal expectation, providing the foundation for interpreting silence as either externally shaped or internally driven.

**Table 1.** Descriptions of class events.

| Event | Event Description |
|---|---|
| Question to directed student | The teacher assigns a specific student to answer a question. |
| Question to all students | The teacher asks a question to the whole class. Each student voluntarily chooses whether or not to answer. |
| New content teaching | The teacher teaches new knowledge, and there is no verbal interaction with the students. |
| Quiz elaboration | The teacher explains the test papers or homework, and there is no verbal interaction with the students. |
| Group activity | Several students form small groups nearby to check each other's recitation of the text or discuss a specific question freely. |
| Quiz | A period of independent practice time without teacher participation, including doing test papers, dictation, and other related activities. |

**2.2. Silence Typology and Behavioral Coding**

Drawing on the event classification above, this section defines and labels students' observable behaviors—particularly silent responses—based on their cognitive and motivational underpinnings. We distinguish between two key types of silence:

- Structural silence refers to silence that is externally expected and regulated by classroom norms. It is common in teacher-centered activities where verbal participation is neither required nor permitted, and aligns with institutionalized expectations of attentiveness and discipline (Fidyk, 2013; Clarke, 2020).
- Strategic silence, by contrast, stems from individual learner decisions. It may reflect contemplation, hesitation, self-protection, or disengagement, and is particularly salient in settings that allow voluntary participation. Prior work links such silence to cognitive uncertainty (Asterhan et al., 2015), identity negotiation (Schultz, 2010), and metacognitive regulation (Biggs, 1996).

This typology allows us to move beyond binary interpretations of silence and instead evaluate its pedagogical and psychological meanings in context. To enable systematic coding of video data, we identify distinct student behaviors within *questioning events* and assign them corresponding status labels (see Table 2). For example:

- In *QtDS* events, when a named student remains silent, the behavior is classified as strategic silence, often indicating a reflective or uncertain stance. If the same student answers, this is driven speaking, prompted by the teacher's directive. Unnamed students who remain silent are in structural silence, while those who voluntarily respond are marked as active speakers.
- In *QtAS* events, where participation is open, silence is interpreted as strategic, and speech is coded based on spontaneity (active speaking), delayed consensus (driven speaking), or topic irrelevance (irrelevant speaking).

**Table 2**. Classification of students' observable behaviors and corresponding silence labels during questioning events

| Event | Student Behavior | Status Label |
| --- | --- | --- |
| Question to directed student | The named student kept silent. | Strategic silence |
| | The named student responds to the question. | Driven speaking |
| | The unnamed student voluntarily responds. | Active speaking |
| | The unnamed students remains silent. | Structural silence |
| | The student speak about unrelated matters. | Irrelevant speaking |
| Question to all students | The student chooses not to respond. | Strategic silence |
| | The student voluntarily answers the question. | Active speaking |
| | The student echoes or agrees with peers after observing others' answers. | Driven speaking |
| | The student speaks about unrelated matters. | Irrelevant speaking |

By establishing a dual-layered framework—first categorizing classroom events and then classifying student behaviors based on the underlying drivers of silence—this study provides a structured and context-sensitive approach to analyzing classroom silence. The distinction between strategic and structural silence, together with the labeling of observable behaviors, enables us to interpret silent and verbal responses in alignment with both theoretical perspectives and real-world instructional dynamics. This framework also ensures compatibility with video-based data, where internal states must be inferred from external cues.

In the following section, we detail how this framework was operationalized in our research through the development of a coding scheme, the annotation of classroom videos, and the subsequent analysis of student silence patterns across different instructional contexts.

## 3. Methods

Based on the framework for classifying student silence during classroom learning described in Section 2, both behavioral and neurophysiological features are collected to investigate different silence patterns. Then, students are divided into three groups based on their academic performance to further explore the potential heterogeneous influences. Figure 1 illustrates the overall design framework of this study. The following section presents the data collection and analysis pipeline.

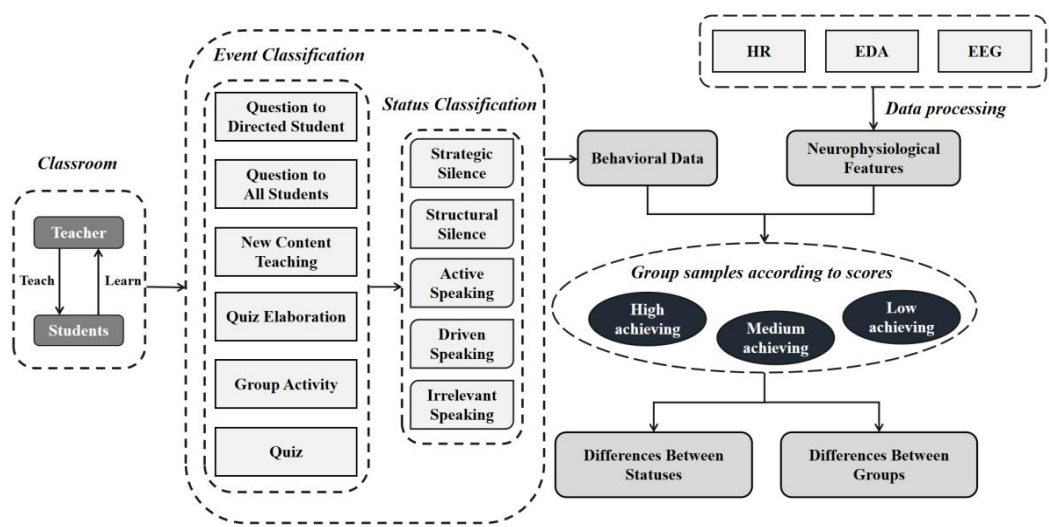

*Figure 1:* The research design framework

### 3.1. Data Collection

To validate the proposed framework, the present study collected multi-modal data from 34 math lessons, covering 54 students (from 2 classes, eighth grade, 13-15 years old) from a middle school. For each 40-min lesson, classroom activities were recorded with two cameras—one in front of the classroom, near the podium and the other at the back door to collect the video and sound. Then, for each student, an EEG headband and a wristband were used to collect their EEG, EDA, and heart rate signals simultaneously, as shown in Fig.2. Specifically, the EEG headband covers students' frontal area (positioned at Fp1 and Fp2) with two dry electrodes.

In addition, students are grouped based on their test score ranking in further analysis. The test scores were the final exam score in the semester before data collection. Following the approach of Sedova and Navratilova (2020), students were divided into three groups: the high-achieving group (top 25%, 13 students), the medium-achieving group (middle 50%, 27 students), and the low-achieving group (bottom 25%, 14 students).

The study adhered to the Declaration of Helsinki guidelines and received approval from the ethics committee of the Department of Psychological and Cognitive Science at Tsinghua University. Written informed consent was obtained from all participants and their legal guardians.

A.

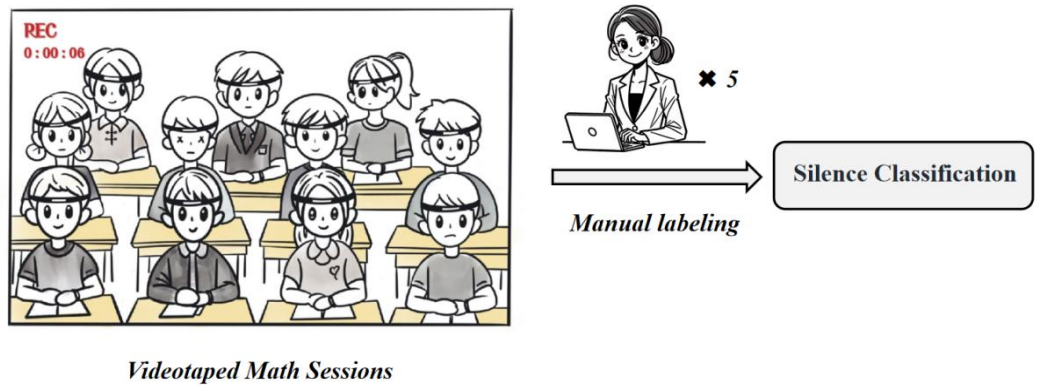

B.

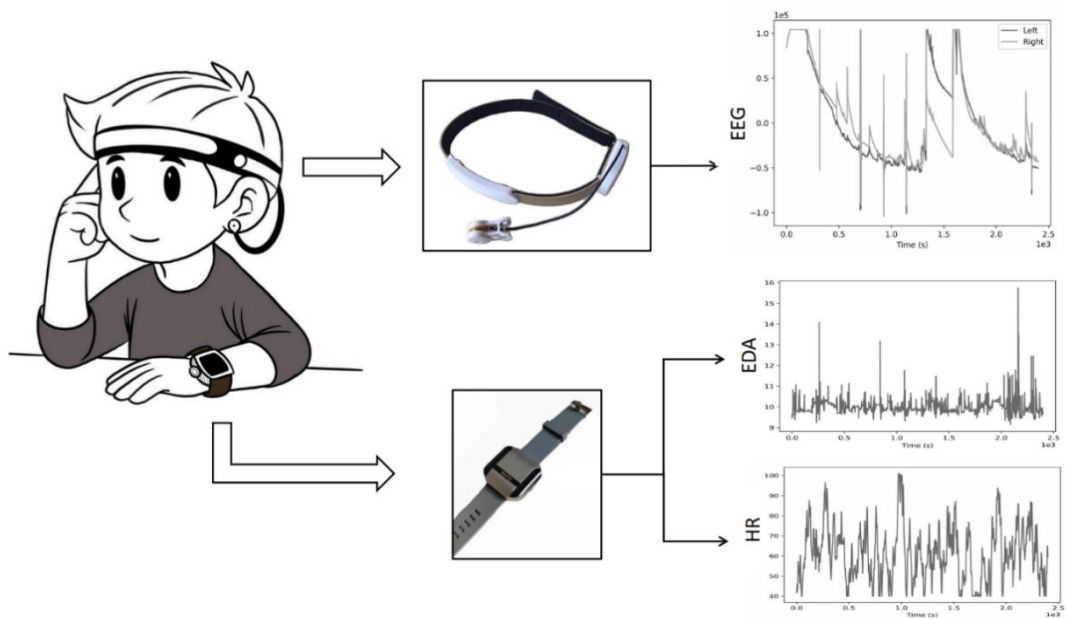

*Figure 2:* The experimental design. **a.** Students wear neurophysiological devices in the classroom. Cameras are used to record class sessions, and student behavioral data is obtained through manual coding. **b.** Wearable neurophysiological devices are displayed. The acquired raw signal samples, including EEG, EDA, and heart rate are plotted on the right. **Note:** The schematic drawing of cameras is assisted by artificial intelligence tools, and the rest of the pictures are hand-painted by the first author.

### 3.2. Video Coding

Class events and students' status in the 34 videos were coded following the framework outlined in Section 2. Five coders participated in the coding, including four graduate students and a middle school teacher with practical teaching experience. All coders joined a training phase before beginning coding formally to master the detail of the framework.

The coding process involves five steps: labeling class events, event labels validation, segmenting videos according to events, labeling student statuses within each event, and status labels validation. Specifically, each class video was independently coded by two individuals using a video playback platform developed by our research group. Coders primarily rely on the videos captured by the camera positioned near the podium. For activities occurring beyond its coverage, the camera at the back door serves as a supplementary perspective. Then, repeated sampling and reliability testing were employed to validate the event labels. If the discrepancy in labeling for two coders exceeded 5% of the total duration (i.e., more than 120 seconds in a 40-minute video), a third individual was assigned to re-evaluate the video. For discrepancies below this threshold, the inconsistent seconds were discarded. Based on the labeling results, the videos were segmented into clips of varying lengths, each representing distinct class events. Notably, during the coding of questioning events, the teacher's question time was excluded, retaining only the student response time. Then, for the status coding, students' status in each event was coded independently by two individuals. In cases of disagreement, a third coder conducted a re-assessment. Video segments where students' status could not be clearly determined due to quality issues were excluded from the analysis. Ultimately, the overall labeling consistency across all videos exceeded 90%.

Based on the results of video coding, the duration of silence was calculated by measuring the total time spent in each category of silence, while the frequency was calculated by the proportion of occurrences of each silence relative to the total sample. These indicators were then used for further analysis of students' behaviors in the classroom.

### 3.3. Neurophysiological Preprocessing and Feature Extraction

The analysis pipeline of EDA, heart rate and EEG signals recorded during class activities were listed as follows.

EDA signals were collected at a sampling rate of 40 Hz. For preprocessing, the signals were segmented into 1-second time windows, with each window containing 40 data points. These segments were then analyzed to identify potential outliers. For each segment, if more than 20% of the data points fall below 0.01, the segment is flagged as missing data. Finally, the data retention rate is 56.26%. The remaining data is decomposed into the slowly changing tonic component and the rapidly changing phasic component (Boucsein, 2012). A high tonic value indicates a higher level of emotional arousal, possibly caused by stress and concentration, and an active status of participation in class (Zhang, 2018). Conversely, a lower tonic level may suggest that the student is experiencing boredom or fatigue (Jang et al., 2015). Similarly, elevated phasic activity, characterized by frequent or pronounced skin conductance responses, indicating heightened sympathetic nervous system activity triggered by stimuli (Posada-Quintero & Chon, 2020). In contrast, diminished phasic activity may be associated with reduced emotional reactivity or decreased attentional engagement during classes (Braithwaite et al., 2013).

At the same time, heart rates were collected at a rate of 1 Hz, with a threshold of 40 bpm applied for marking outliers. A total of 94.38% of the heart rate data was retained. Heart rates were expected to help reflect whether students are in a status of tension or relaxation (Benson et al., 1974).

For the EEG data, the relative power spectral density (RPSD) of $\delta$ (0.5-4Hz), $\theta$ (4-7Hz), $\alpha$ (8-12Hz), $\beta_{low}$ (13-18Hz), $\beta_{high}$ (18-30Hz), and $\gamma$ (> 30Hz) bands were calculated for each 1-s

EEG data, after marking missing segments, detrending, eliminating artifacts and oculograms, filtering (0.5-50Hz), and segmenting (Xu et al., 2024). RPSD serves as a valuable indicator for understanding fluctuations in cognitive and emotional processes during learning (Klimesch, 1999). By normalizing power across frequency bands, RPSD minimizes the influence of inter-individual variability in overall EEG amplitude (Pivik et al., 1993). Different frequency bands in EEG signals correspond to distinct cognitive functions. Lower-frequency bands, such as $\alpha$ waves, are associated with relaxed states (Zoefel et al., 2011; Klimesch, 1999). In contrast, higher-frequency bands, such as $\beta$ and $\gamma$ waves, are linked to advanced cognitive functions, including information integration and problem-solving (Doyle et al., 1974).

Finally, student status data derived from videos and the corresponding neurophysiological data were aligned on a scale of 1-second, with valid data for statistical analysis requiring both concurrent status labels and neurophysiological values for each student. Considering the varying durations of events, the mean value of neurophysiological data within each event is calculated to characterize students' implicit status.

### 3.4. Statistical Analyses

An Ordinary Least Squares (OLS) regression model was proposed to investigate the relationship between academic achievement and different types of silence. The frequency of each type of silence was calculated as the dependent variable, and test score as the independent variable.

$$Freq_i = \gamma_0 + \gamma_1 Score_i + \eta_i \#(1)$$

where $Freq_i$ denotes the frequency of the specific status (i.e., structural silence in QtAS and QtDS, strategic silence and active speaking in QtAS) for $student\ i$, $Score_i$ is the final exam score of student $i$. $\gamma_0$ is the intercept, $\gamma_1$ denotes the coefficient of score, and $\eta_i$ is the random error term.

Furthermore, to explore the possible implicit differences during silence, the present study employed a linear mixed-effects (LME) model to compare neurophysiological features across different learning statuses. Event-status are treated as independent variables, with neurophysiological features as dependent variables. To address the potential intraclass correlation (ICC) of statuses within individual students, student IDs were included as categorical variables to group the observations accordingly. The model is as follow:

$$Feature_{ij} = \beta_0 + \beta_1 Status_{ij} + u_j + \epsilon_{ij}\#(2)$$

where $Feature_{ij}$ represents the value of the neurophysiological feature for student $i$ in the specific status $j$. $Status_{ij}$ denotes the status $i$ of the student $j$. $\beta_0$ is the intercept, while $\beta_1$ is the regression coefficient for the status categorical variable. $u_j$ is the random effect associated with student $j$, accounting for the variability between students, and $\epsilon_{ij}$ is the residual error term, capturing the variability within students. Pairwise paired post hoc F-tests are conducted on $\beta_1$ to assess whether the differences between coefficients are statistically significant.

Moreover, in order to explore the possible effect of individual difference in academic achievement, another LME model is further employed to investigate the differences neurophysiological features across various academic performance during the same status. For a given status, achievement groups are treated as independent variables, and neurophysiological features serve as the dependent variables. The model is as follow:

$$Feature_{mk} = \beta_2 + \beta_3 Group_{mk} + u_k + \epsilon_{mk} \#(3)$$

where $Feature_{mk}$ represents the neurophysiological value of the student $k$ in the specific academic performance group $m$. $Group_{mk}$ denotes the student $k$ in the academic group $m$. $\beta_2$ is the intercept, and $\beta_3$ is the regression coefficient of the group categorical variable. $u_k$ is the random effect of the student $k$, representing the variability between students, and $\epsilon_{mk}$ is the residual error term, accounting for within-student variability. Pairwise paired post hoc F-tests are conducted on $\beta_3$ to assess the statistical significance of the observed differences of features.

## 4. Results

### 4.1. Sample Description

Table 3 presented the duration of each type of event and student status within each event. The questioning events (i.e., QtDS and QtAS) occupied 50.5% of the relative duration of lessons. However, in these questioning events, silence constituted a significant portion of the overall response. During QtDS, structural silence accounted for nearly 85% of the total duration, while strategic silence represented 1.83%. In addition, when the teacher asked a question to the entire class (i.e., QtAS), students engaged in strategic silence for 78.5% of the duration. Moreover, in the unidirectional teaching events (i.e., new content teaching and quiz elaboration), whose proportion was 37.16%, all students maintained structural silence throughout this period. In the questioning sessions, students also exhibited active speaking behaviors, accounting for 2.31% in QtDS and 10.61% in QtAS. Quiz and group activities occurred infrequently in lessons, comprising only 11.92% and 0.42% of the total class time, respectively. During group activities, all students were considered in the active speaking status.

**Table 3:** The relative duration proportions of events and status in the specific event in 34 math lessons.

| Event | Proportion of event | Status | Proportion of status in specific event |
|---|---|---|---|
| Question to directed student | 11.72% | Structural silence | 84.20% |
| | | Strategic silence | 1.83% |
| | | Active speaking | 2.31% |
| | | Driven speaking | 5.69% |
| | | Irrelevant speaking | 5.97% |
| Question to all students | 38.78% | Strategic silence | 78.50% |
| | | Active speaking | 10.61% |
| | | Driven speaking | 5.92% |
| | | Irrelevant speaking | 4.98% |
| New content teaching | 16.09% | Structural silence | 100% |
| Quiz elaboration | 21.07% | Structural silence | 100% |
| Group activity | 0.42% | Active speaking | 100% |
| Quiz | 11.92% | — | — |

Figure 3 further demonstrated the differences on students' status across groups with various academic performance. Different status patterns could be observed across different groups: the high-achieving groups showed the lowest strategic silence and the most active speaking during lessons, compared with the medium-achieving and low-achieving groups; the medium-achieving students had a higher occurrence of strategic silence (QtAS) than that of the low-achieving group and the high-achieving group. Although all groups showed few strategic silence (QtDS) during their real classroom activities, the low-achieving group, on the other hand, had more occurrences for the specific status.

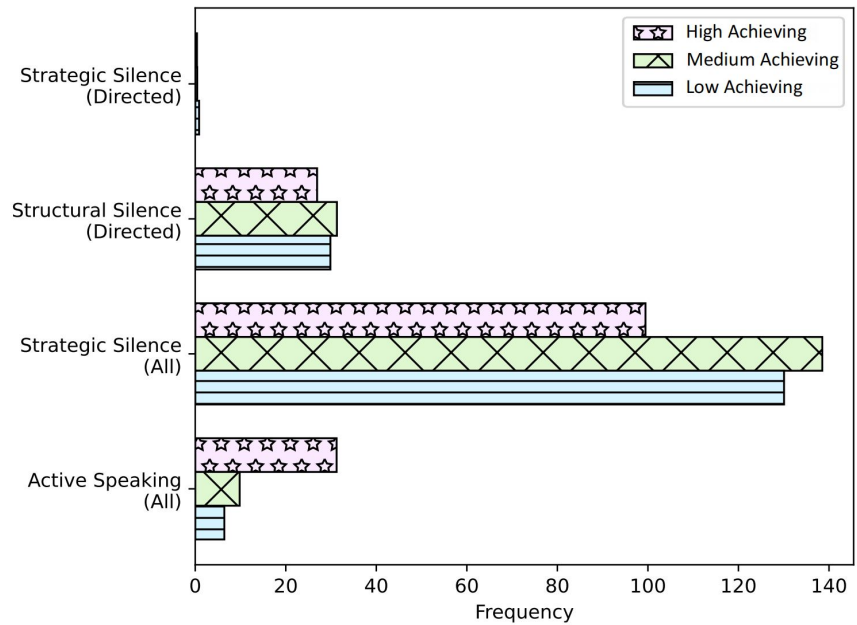

*Figure 3:* The frequency of four status patterns (three types of silence and actively speaking) by three achievement groups.

## 4.2. The Relationship between Silence and Academic Performance

Figure 4 exhibited the relationship between academic performance and frequency of specific type of status respectively. Specifically, a significant positive correlation between academic performance and the frequency of students taking the initiative to speak during QtAS was found, indicating that students with higher achievement were more inclined to speak up ($r = 0.408$, $p = .006$). Conversely, significant negative correlations between academic performance and the frequencies of three types of silence were found (strategic silence during QtAS: $r = -.479$, $p = .001$; strategic silence during QtDS: $r = -.617$, $p = .005$; structural silence in QtDS: $r = -.424$, $p = .005$), suggesting that more silence may be associated with lower achievement. However, this trend did not persist within subgroups. Specifically, among high-achieving students, no significant correlation was found between silence and test scores (strategic silence in QtDS: $r = -.046$, $p = .115$; structural silence in QtDS: $r = -.335$, $p = .454$; strategic silence in QtAS: $r = -1.1$, $p = .104$). Among medium-achieving students, a negative correlation emerged between strategic silence in QtAS and test scores ($r = -.377$, $p = .032$), whereas no significant associations were observed in other dimensions (strategic silence in QtDS: *r* = -.013, *p* = .467; structural silence in QtDS: *r* = -.203, *p* = .103). Similarly, no significant relationships were identified in the low-achieving group (strategic silence in QtDS: *r* = -.032, *p* = .342; structural silence in QtDS: *r* = -.061, *p* = .514; strategic silence in QtAS: *r* = -.121, *p* = .467).

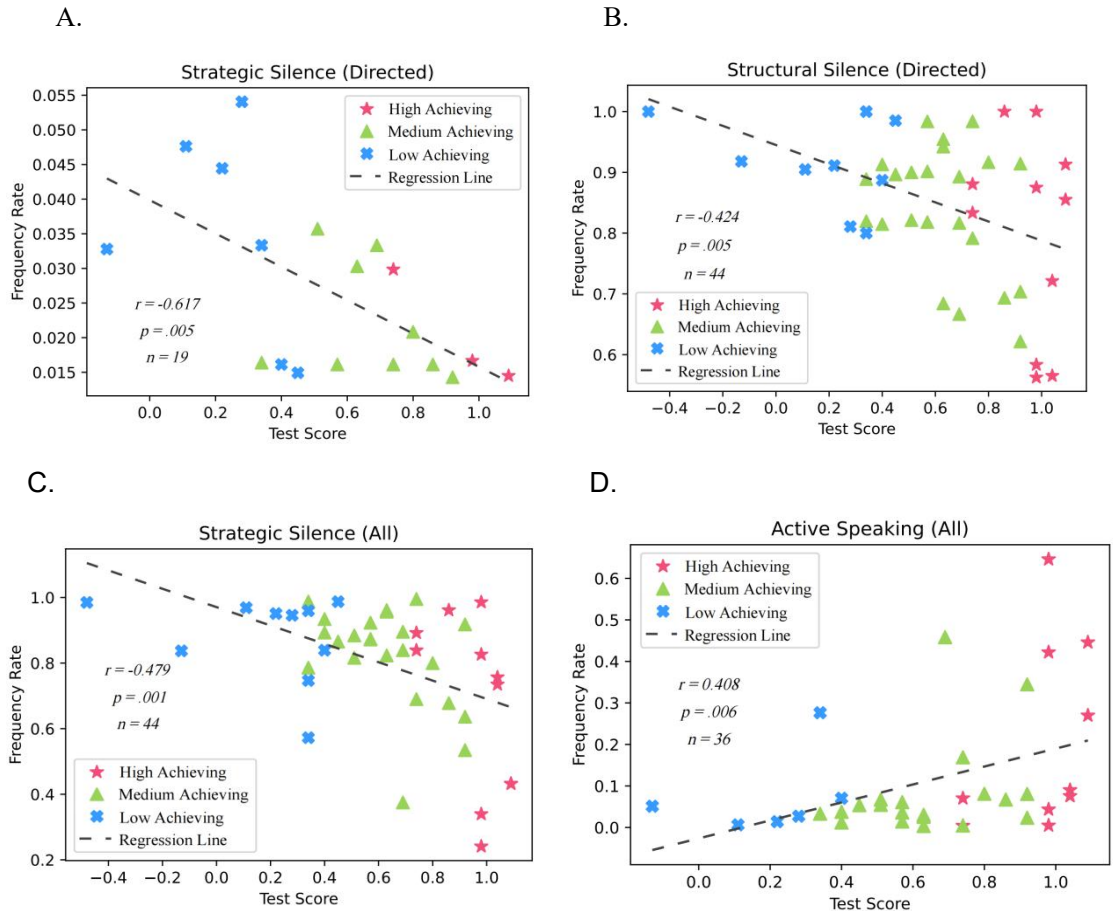

*Figure 4:* Scatter plots between academic performance and different learning status. The horizontal axis represents the final exam scores, and the vertical axis denotes the frequency of the specific status, as detailed in the subheadings.

### 4.3. Neurophysiological Representations of Silence

Table 4 demonstrated the differences of neurophysiological representations across different learning status, including heart rates and EDA components, on the entire sample. Specifically, students exhibited substantially higher heart rates and the tonic part of EDA in structural silence between QtDS compared to the same type of silence in teaching (both *p* < 0.01). Within the category of questioning events, students exhibited higher values in heart rates and EDA tonic component in structural silence in QtDS than strategic silence during QtAS session (both *p* < 0.01). Nevertheless, for strategic silence in both QtDS and QtAS, no differences in heart rates and EDA indicators were found. Also, for both structural silence and strategic silence in QtDS, there was no significant difference in those representations.

For EEG in Table B1, students showed higher RPSD of $\gamma$ band in structural silence during QtDS compared to both the same type of silence in teaching and strategic silence in QtAS (both *p*

< 0.05). Other comparisons of RPSD did not reach statistical significance.

Importantly, during QtAS, students showed lower heart rates in strategic silence compared to active speaking ($p < 0.01$), while there was no significant difference for EDA tonic and phasic values between the two statuses. During structural silence in QtDS, students exhibited notably higher heart rates and lower EDA tonic component compared to active speaking in the group activity (heart rate: $p < 0.01$; EDA tonic: $p < 0.05$). There were no significant differences in heart rates and EDA indicators between structural silence during teaching and speaking during group activity.

**Table 4:** Comparison of heart rates and EDA features between different statuses based on the entire sample.

| Status 1 | Status 2 | Mean Difference (Status1 - Status2) | | |
|---|---|---|---|---|
| | | **Heart Rate** | **EDA tonic** | **EDA phasic** |
| Strategic Silence (All) | Active Speaking (All) | -4.213** (1.470) | 0.069 (0.130) | 0.007 (0.003) |
| Structural Silence (Directed) | Active speaking (Group Activity) | 4.198** (1.543) | -0.267* (0.136) | 0.008 (0.004) |
| Structural Silence (Teaching) | Active speaking (Group Activity) | 1.162 (1.447) | -0.477 (0.131) | 0.007 (0.003) |
| Structural Silence (Directed) | Structural Silence (Teaching) | 3.037** (0.609) | 0.210** (0.047) | 0.002 (0.003) |
| Strategic Silence (Directed) | Strategic Silence (All) | -3.730 (2.690) | 0.614 (0.465) | -0.013 (0.001) |
| Structural Silence (Directed) | Strategic Silence (Directed) | 5.893 (2.737) | -0.385 (0.466) | 0.017 (0.002) |
| Structural Silence (Directed) | Strategic Silence (All) | 2.162** (0.637) | 0.158** (0.049) | 0.004 (0.002) |

Notes: **$p < 0.01$, *$p < 0.05$. Standard errors in parentheses.

Table 5 further exhibited the differences of neurophysiological representations across various academic achievements. For the medium-achieving group, mean heart rates during strategic silence were much lower than that during active speaking, consistent with the entire sample in Table 4 ($p < 0.01$). Nevertheless, the trend was not found in the results of other groups, which revealed the necessity of using academic achievement to group individuals. High-achieving students indicated significant higher heart rates in the structural silence during QtDS and new content teaching compared to active speaking during group activity ($p < 0.01$). Low-achieving group had lower EDA tonic values in the structural silence during QtDS than active speaking during group activity (p < 0.05). For all groups, students demonstrated lower EDA tonic values in the structural silence during new content teaching than active speaking during group activity ($p < 0.05$). In the situations above, EDA phasic components did not show any difference in all groups ($p > 0.05$).

Moreover, compared to structural silence during teaching sessions, in the corresponding silence observed in QtDS, high-achieving students displayed lower EDA phasic values ($p < 0.05$); low-achieving students showed no difference in EDA tonic component, which differed from results presented in Table 4 ($p > 0.05$). High-achieving students exhibited higher heart rates and EDA tonic during structural silence in QtDS compared to strategic silence during QtAS ($p < 0.01$). For high-achieving and low-achieving students, higher heart rates were also observed during structural silence in QtDS compared to strategic silence in QtAS (both $p < 0.01$). For high-achieving and medium-achieving students, the EDA tonic during structural silence in QtDS was higher than during strategic silence in QtAS (high-achieving: $p < 0.01$; medium-achieving: $p < 0.05$).

**Table 5:** Comparison of heart rates and EDA features among different statuses by student achievement level.

| Status 1 | Status 2 | Mean Difference (Status1 - Status2) | | | | | | | | |
|---|---|---|---|---|---|---|---|---|---|---|
| | | Heart Rate | | | EDA tonic | | | EDA phasic | | |
| | | High | Medium | Low | High | Medium | Low | High | Medium | Low |
| Strategic Silence (All) | Active Speaking (All) | 1.405 (1.913) | -8.071** (2.314) | -3.263 (3.376) | 0.119 (0.178) | -0.068 (0.192) | 0.041 (0.366) | 0.005 (0.002) | 0.003 (0.005) | 0.016 (0.005) |
| Structural Silence (Directed) | Active Speaking (Group Activity) | 8.868** (2.556) | 2.830 (2.224) | 1.403 (3.513) | -0.290 (0.259) | -0.142 (0.177) | -0.624* (0.369) | 0.002 (0.005) | 0.012 (0.005) | 0.005 (0.013) |
| Structural Silence (Teaching) | Active Speaking (Group Activity) | 5.243* (2.364) | 0.616 (2.105) | -2.241 (3.256) | -0.614* (0.248) | -0.406* (0.170) | -0.617* (0.352) | 0.018 (0.006) | 0.005 (0.003) | -0.007 (0.012) |
| Structural Silence (Directed) | Structural Silence (Teaching) | 3.625** (1.132) | 2.273** (0.829) | 3.644** (1.414) | 0.248* (0.090) | 0.264** (0.060) | -0.056 (0.124) | -0.016* (0.005) | 0.007 (0.003) | 0.012 (0.005) |
| Strategic Silence (Directed) | Strategic Silence (All) | -5.972 (1.853) | -5.873 (3.908) | 12.854 (0.662) | 0.098 (0.335) | 1.051 (0.799) | — | -0.007 (0.002) | -0.014 (0.002) | — |
| Structural Silence (Directed) | Strategic Silence (Directed) | 7.880 (2.058) | 6.869 (3.970) | -6.432 (1.367) | 0.247 (0.345) | -0.918 (0.801) | — | 0.010 (0.002) | 0.019 (0.003) | — |
| Structural Silence (Directed) | Strategic Silence (All) | 1.908** (1.194) | 0.997 (0.848) | 6.422** (1.519) | 0.346** (0.083) | 0.133* (0.061) | 0.032 (0.131) | 0.003 (0.003) | 0.006 (0.004) | 0.002 (0.007) |

Notes: High, medium, and low are abbreviations for high-achieving, medium-achieving, and low-achieving respectively. **$p < 0.01$, *$p < 0.05$. Standard errors in parentheses.

The EEG results from the grouped sample, as presented in Table 6, provided further evidence on the distinct neurophysiological patterns among different types of silence and further demonstrated the influence of group heterogeneity. High-achieving students had higher RPSD of α and $\beta_{low}$ band for structural silence during QtDS than that during the teaching event (α band: $p < 0.01$; $\beta_{low}$ band: $p < 0.05$). No significant differences in RPSD values were observed across the various frequency bands for all groups when comparing strategic silence during QtDS and QtAS, as well as structural silence and strategic silence during QtDS (all $p > 0.05$). High-achieving students displayed lower RPSD of $\gamma$ band in the structural silence during QtDS, in comparison of strategic silence during QtAS ($p < 0.01$).

At the same time, high-achieving students demonstrated higher RPSD of $\beta_{high}$ and $\gamma$ bands during strategic silence in QtAS, compared to active speaking in the same event (both $p < 0.05$). For the three groups with varying academic performance, structural silence during QtDS and teaching showed no significant differences across all frequency bands, compared to active speaking in group activity. Additionally, both the high-achieving and low-achieving groups showed lower RPSD of θ band in the same comparison (both $p < 0.05$).

The further comparisons across different academic achievement groups were detailed provided in Appendix C, demonstrating the group heterogeneity during silence.

**Table 6:** Comparison of EEG features (RPSD) of various frequency bands between different statuses by student achievement level.

| Status 1 | Status 2 | Mean Difference (Status1 - Status2) | | | | |
|---|---|---|---|---|---|---|
| | | EEG θ | EEG α | EEG $\beta_{low}$ | EEG $\beta_{high}$ | EEG γ |

| | | High | Medium | Low | High | Medium | Low | High | Medium | Low | High | Medium | Low | High | Medium | Low |
|---|---|---|---|---|---|---|---|---|---|---|---|---|---|---|---|---|
| Strategic Silence (All) | Active Speaking (All) | -0.010* (0.011) | -0.008 (0.006) | -0.015* (0.011) | 0.015 (0.005) | -0.003 (0.004) | 0.000 (0.005) | 0.017 0.003 | -0.002 (0.003) | -0.001 (0.003) | 0.027* (0.006) | 0.001 (0.005) | -0.016 (0.010) | 0.012* (0.002) | -0.001 (0.002) | -0.010 (0.005) |
| Structural Silence (Directed) | Active Speaking (Group Activity) | -0.017 (0.010) | -0.005 (0.005) | -0.004 (0.007) | 0.000 (0.006) | -0.003 (0.003) | -0.004 (0.004) | 0.003 (0.003) | -0.003 (0.003) | -0.003 (0.003) | 0.000 (0.009) | -0.004 (0.007) | 0.003 (0.004) | -0.005 (0.005) | -0.001 (0.003) | 0.002 (0.002) |
| Structural Silence (Teaching) | Active Speaking (Group Activity) | -0.018 (0.010) | -0.005 (0.005) | -0.002 (0.006) | -0.014 (0.005) | -0.003 (0.003) | -0.001 (0.004) | -0.007 (0.002) | -0.002 (0.003) | -0.002 (0.003) | -0.008 (0.008) | 0.000 (0.006) | 0.002 (0.004) | -0.005 (0.005) | 0.002 (0.003) | 0.002 (0.001) |
| Structural Silence (Directed) | Structural Silence (Teaching) | 0.001 (0.003) | 0.000 (0.003) | -0.002 (0.004) | 0.014** (0.003) | -0.001 (0.001) | -0.003 (0.002) | 0.010* (0.002) | -0.001 (0.001) | -0.002 (0.001) | 0.008 (0.003) | -0.004 (0.002) | 0.001 (0.002) | 0.000 (0.002) | -0.002 (0.001) | 0.000 (0.001) |
| Strategic Silence (Directed) | Strategic Silence (All) | -0.002 (0.003) | 0.002 (0.008) | -0.005 (0.003) | 0.003 (0.010) | 0.008 (0.006) | 0.004 (0.003) | 0.011 (0.008) | 0.001 (0.003) | 0.004 (0.003) | 0.025 (0.013) | 0.001 (0.007) | -0.001 (0.004) | 0.005 (0.003) | 0.001 (0.002) | -0.001 (0.002) |
| Structural Silence (Directed) | Strategic Silence (Directed) | 0.003 (0.003) | 0.000 (0.008) | 0.008 (0.005) | -0.005 (0.010) | -0.007 (0.006) | -0.003 (0.003) | -0.014 (0.008) | -0.001 (0.003) | -0.003 (0.003) | -0.034 (0.013) | 0.001 (0.007) | 0.003 (0.004) | -0.011 (0.003) | 0.000 (0.002) | 0.000 (0.002) |
| Structural Silence (Directed) | Strategic Silence (All) | 0.000 (0.003) | 0.001 (0.003) | 0.003 (0.004) | -0.002 (0.003) | 0.000 (0.001) | 0.001 (0.002) | -0.003 (0.002) | 0.001 (0.001) | 0.001 (0.001) | -0.009 (0.004) | 0.002 (0.002) | 0.002 (0.002) | -0.006** (0.002) | 0.001 (0.001) | 0.000 (0.001) |

Notes: High, medium, and low are abbreviations for high-achieving, medium-achieving, and low-achieving respectively. **$p < 0.01$, *$p < 0.05$. Standard errors in parentheses.

## 5. Conclusion and Discussion

This study aimed to explore the complex and often uncovered phenomenon of classroom silence through an integrative, multi-modal approach. We incorporated both behavioral data (i.e., classroom video coding) and neurophysiological data (i.e., EEG, EDA, and heart rate) to examine different types of silence. Grounded in the theoretical distinction between *event* and *status*, we proposed a novel classification framework that differentiates between structural and strategic silence in distinct classroom contexts. Our analysis revealed how silence varies across student achievement levels, offering new insights into the emotional and cognitive underpinnings of silent classroom behavior.

Through behavioral indicators based on video coding, we identified distinct types of silence that were influenced by classroom context, partially supporting the rationality of our proposed framework. These findings align with previous studies that highlight the roles of situational factors and individual traits in shaping silence (Bao, 2020). Although prior literature has discussed forms of structural and strategic silence (Jaworski and Sachdev, 1998; Bassett, 2014; Bosacki, 2005), our study advances this understanding by operationalizing silence through an event-status lens and verifying it with multi-modal evidence. We captured silence as a dynamic and heterogenous phenomenon, rather than a monolithic behavioral marker.

The neurophysiological evidence provided further support for this classification. For instance, comparative analysis reveals higher heart rates and EDA tonic components during structural silence in QtDS contexts, relative to both structural silence in new content teaching and strategic silence in QtAS. These markers are typically associated with cognitive-emotional status of anticipatory anxiety and attentional strain (Setz et al., 2009). Therefore, this pattern suggests that the prospect of being individually addressed during QtDS may elicit increased cognitive load and evaluative apprehension (Ayres et al., 2021). Such differential activations support the necessity of silence classification, revealing the complexity of students' silence.

Additionally, compared with active speaking status, which is often regarded as positive for learning, high-achieving students displayed no significant differences in heart rates and both EDA components compared with those during the strategic silence of QtAS. In contrast, a higher EEG high-frequency RPSD during these moments was found. Increased high-frequency RPSD indicates heightened cognitive load and intensive information processing (Constant and Sabourdin, 2012; Castro-Meneses et al., 2020). These findings suggest that high-achieving students may remain cognitively engaged despite behavioral inactivity. These findings challenge the simplistic association of silence with disengagement, emphasizing the need to reconsider silence not just as the absence of speech but as a potential marker of internal processing.

While some prior research has acknowledged the constructive roles of silence (e.g., Fidyk, 2013; Tang et al., 2020), empirical evidence supporting this perspective has been limited. Our results fill this gap by demonstrating that certain silence types, particularly strategic silence among high-achieving students, are accompanied by heightened cognitive activity. Thus, silence is not inherently passive or active—it reflects a spectrum of internal states shaped by context and individual attributes.

Furthermore, the relationship between silence and academic achievement revealed meaningful patterns. Low-achieving students exhibited higher frequencies of undifferentiated silence across contexts, potentially reflecting anxiety, lack of understanding, or fear of peer judgment (Meyer & Turner, 2006; Liu & Jackson, 2008). In contrast, high-achieving students used silence more strategically, as evidenced by neurophysiological indicators of sustained attention and reflection. These findings indicate that silence should not be uniformly interpreted as disengagement, but rather understood within the broader context of student goals, classroom dynamics, and individual learning strategies.

Neurophysiological heterogeneity across achievement groups further underscores the complexity of silent behavior. For example, medium-achieving students displayed reduced heart rates during strategic silence in QtAS compared to active speaking within the same event. This suggests that average-achieving students may experience a shift toward a more relaxed status during strategic silence (Asterhan et al., 2015), which, however, did not hold among high-achieving students and low-achieving students. Regarding structural silence across different events, high-achieving students were found to display higher RPSD values in the γ brand during strategic silence in QtAS compared to the structural silence in QtDS. The elevated RPSD signature implies that high-achieving students are more cognitively engaged during strategic silence. The absence of differentiated neurophysiological responses among low-achieving students between structural silence in QtDS and other events points to potential difficulties in adaptive classroom engagement (Ollin, 2008; Setz et al., 2009; Ayres et al., 2021). Such group-based differences highlight the value of combining neurophysiological data with behavioral observation to reveal the latent dimensions of classroom participation.

Importantly, this study demonstrated the unique strengths of a multi-modal approach. EEG, EDA, and heart rate data each capture different aspects of internal states—cognitive processing, emotional arousal, and physiological stress regulation, respectively—providing a more comprehensive picture of what silence *means* in real-time learning. This divergence among modalities also implies that silent behaviors are shaped by asynchronous processes, further validating the necessity of an integrated analytical approach.

Our work contributes to the theoretical and methodological development of silence research in education. It is the first to longitudinally examine classroom silence using both behavioral and physiological data in naturalistic settings. Unlike prior studies that largely relied on post-hoc reflection or single-modality observations, this study provides real-time, multi-dimensional evidence of how silence operates within authentic learning environments. By reconceptualizing silence as a contextually embedded and cognitively meaningful phenomenon, we move beyond traditional assumptions and pave the way for more inclusive and flexible pedagogical strategies.

Despite its contributions, this study has several limitations. First, its findings are based on data collected solely from Chinese classroom contexts, which may limit cross-cultural generalizability. Future research should include more culturally diverse settings to assess the robustness of our framework. Second, while academic achievement served as a proxy for individual differences, it does not fully account for learner variability. Future studies should incorporate factors such as gender, personality, or cultural background (Kovalainen & Kumpulainen, 2007) to enrich our understanding of silence as a socially and individually constructed phenomenon.

## Acknowledgement

This work was supported by the National Natural Science Foundation of China (NSFC) (62177030 and 62477027).

## Appendix A. The labeling ratio of individual status

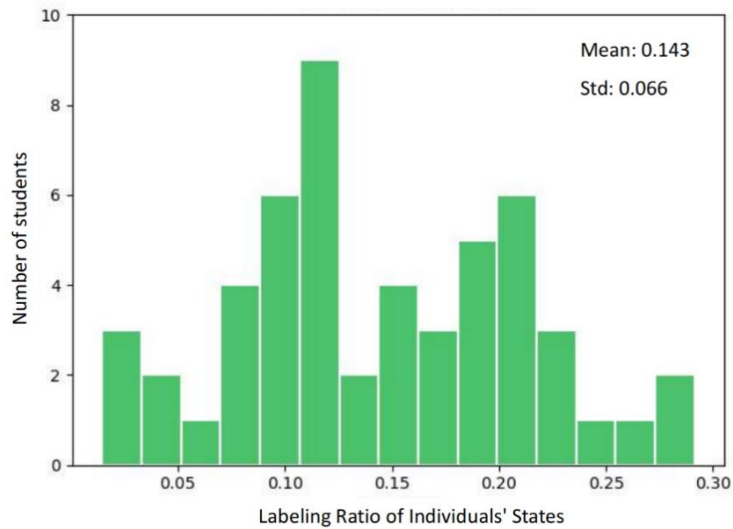

*Figure A1:* The distribution of labeling ratios for individual behavioral status. When calculating this ratio, the denominator represents the total questioning duration across 34 classes, while the numerator reflects the length of time that an individual annotated with a specific status. The average labeling ratio for the entire sample is 14.3%, with a variance of 0.066.

## Appendix B. Results of EEG representations of the entire sample

**Table B1:** Comparison of EEG features (RPSD) between different statuses based on the entire sample.

| Status 1 | Status 2 | Mean Difference (Status1 - Status2) |
| --- | --- | --- |

|  |  | EEG θ | EEG α | EEG $β_{low}$ | EEG $β_{high}$ | EEG γ |
|---|---|---|---|---|---|---|
| Strategic Silence (All) | Active Speaking (All) | -0.010** (0.005) | 0.002 (0.003) | 0.002 (0.002) | 0.000 (0.004) | 0.000 (0.002) |
| Structural Silence (Directed) | Active speaking (Group Activity) | -0.007 (0.004) | -0.002 (0.002) | -0.001 (0.002) | 0.000 (0.004) | 0.000 (0.002) |
| Structural Silence (Teaching) | Active speaking (Group Activity) | -0.007 (0.004) | -0.004 (0.002) | -0.003 (0.002) | 0.000 (0.004) | 0.000 (0.002) |
| Structural Silence (Directed) | Structural Silence (Teaching) | 0.000 (0.002) | 0.002 (0.001) | 0.002 (0.001) | 0.000 (0.001) | 0.000* (0.001) |
| Strategic Silence (Directed) | Strategic Silence (All) | -0.003 (0.005) | 0.005 (0.006) | 0.005 (0.005) | 0.005 (0.008) | 0.001 (0.003) |
| Structural Silence (Directed) | Strategic Silence (Directed) | 0.004 (0.005) | -0.004 (0.006) | -0.003 (0.005) | -0.003 (0.009) | -0.002 (0.003) |
| Structural Silence (Directed) | Strategic Silence (All) | 0.002 (0.002) | 0.001 (0.001) | 0.001 (0.001) | 0.001 (0.002) | 0.000* (0.001) |

**Appendix C. Results of neurophysiological representations by achievement level**

Based on different academic achievement, students' learning statuses under the same context often exhibited group heterogeneity. In Table C1, the low-achieving students demonstrated significantly distinct mean heart rates compared to their peers across all learning statuses. Notably, during strategic silence in QtDS and active speaking in the group activity, heart rates of the low-achieving group were elevated compared to the high-achieving and medium-achieving groups (all $p < 0.01$). In contrast, under other conditions (i.e., structural silence in QtDS and teaching sessions, strategic silence and active speaking in response to QtAS), the low-achieving students' heart rates were lower compared to the other groups, suggesting different levels of tension (all $p < 0.01$). Additionally, for structural silence in QtDS, low-achieving students exhibited lower EDA tonic values than those of the high-achieving and medium-achieving students ($p < 0.05$). During strategic silence in QtAS, the EDA tonic mean for the medium-achieving group was notably higher than the other groups (both $p < 0.01$). These differences in Table C1 further emphasized the distinct neurophysiological responses among student groups.

**Table C1:** Comparison of neurophysiological features between different groups within the specific status.

| Status | Heart Rate | | | EDA tonic | | | EDA phasic | | |
|---|---|---|---|---|---|---|---|---|---|
|  | H.-M. | H.-L. | M.-L. | H.-M. | H.-L. | M.-L. | H.-M. | H.-L. | M.-L. |
| Strategic Silence (Directed) | 0.596 (4.276) | -11.608** (1.767) | -12.204** (3.893) | -1.142 (0.866) | — | — | 0.000 (0.001) | — | — |
| Structural Silence (Directed) | 1.606 (1.310) | 2.704** (1.727) | 1.098** (1.572) | 0.023 (0.100) | 0.319* (0.144) | 0.296* (0.130) | -0.010 (0.004) | -0.010 (0.006) | 0.000 (0.006) |
| Strategic Silence (All) | 0.695 (0.656) | 7.217** (0.867) | 6.523** (0.746) | -0.190** (0.028) | 0.005 (0.056) | 0.195** (0.062) | -0.007 (0.002) | -0.010 (0.005) | -0.003 (0.005) |
| Active Speaking (All) | -8.781 (2.930) | 2.550** (3.782) | 11.331** (4.024) | -0.377 (0.260) | 0.299 (0.403) | 0.676 (0.408) | -0.009 (0.005) | 0.001 (0.001) | 0.010 (0.005) |
| Structural Silence (Teaching) | 0.255 (0.503) | 2.723** (0.546) | 2.468** (0.465) | 0.039 (0.044) | 0.015 (0.053) | -0.024 (0.047) | 0.013 (0.004) | 0.018 (0.004) | 0.005 (0.002) |
| Active Speaking (Group Activity) | -4.431 (3.125) | -4.761** (3.986) | -0.330** (3.849) | 0.171 (0.298) | -0.015 (0.428) | -0.186 (0.388) | 0.000 (0.005) | -0.007 (0.013) | -0.007 (0.012) |

Note: H., M., and L. are abbreviations for high-achieving, medium-achieving, and low-achieving respectively. **$p < 0.01$, *$p < 0.05$.

Table C2 presented the detailed results of EEG cross-group comparisons. During the strategic silence in QtDS, the low-achieving group displayed notable difference in the θ band of RPSD ($p <$

0.05); high-achieving students had higher RPSD values in $\beta_{high}$ and $\gamma$ bands than those of medium-achieving students ($\beta_{high}$: $p < 0.05$; $\gamma$: $p < 0.01$). During other learning statuses listed in the table, low-achieving students showed significant differences in the θ and α bands of RPSD compared to the other groups (all $p < 0.01$). For structural silence in QtDS and strategic silence in QtAS, low-achieving students demonstrated lower RPSD values in $\beta_{low}$ and $\beta_{high}$ bands (all $p < 0.01$); the medium-achieving group showed substantial higher RPSD values in the $\gamma$ band than those of the low-achieving group (structural silence in QtDS: $p < 0.01$; strategic silence in QtAS: $p < 0.05$). For active speaking in QtAS, medium-achieving students had higher RPSD values in the $\beta_{high}$ and $\gamma$ band than those of the low-achieving group (both $p < 0.05$). For both structural silence in teaching events and active speaking in the group activity, low-achieving students showed significant lower $\beta_{low}$ values than other groups (all $p < 0.01$); high-achieving students exhibited higher $\gamma$ values compared to medium-achieving students (both $p < 0.05$). Additionally, for structural silence during teaching, medium-achieving students showed higher RPSD of $\beta_{high}$ band than the low-achieving group ($p < 0.01$). These findings further highlighted the group heterogeneity in cognitive activities, with variations observed based on academic achievement levels.

**Table C2:** Comparison of EEG features (RPSD) of various frequency bands between different groups within the specific status.

| Status | EEG θ | | | EEG α | | | EEG $\beta_{low}$ | | | EEG $\beta_{high}$ | | | EEG γ | | |
|---|---|---|---|---|---|---|---|---|---|---|---|---|---|---|---|
| | H.-M. | H.-L. | M.-L. | H.-M. | H.-L. | M.-L. | H.-M. | H.-L. | M.-L. | H.-M. | H.-L. | M.-L. | H.-M. | H.-L. | M.-L. |
| Strategic Silence (Directed) | -0.010 (0.008) | 0.002* (0.003) | 0.012* (0.008) | 0.009 (0.012) | 0.016 (0.010) | 0.007 (0.007) | 0.022 (0.009) | 0.022 (0.009) | 0.000 (0.004) | 0.053* (0.015) | 0.058 (0.014) | 0.005 (0.008) | 0.019** (0.003) | 0.021 (0.003) | 0.002 (0.003) |
| Structural Silence (Directed) | -0.007 (0.003) | -0.004** (0.004) | 0.004** (0.004) | 0.011 (0.003) | 0.013** (0.003) | 0.003** (0.002) | 0.009 (0.002) | 0.011** (0.002) | 0.002** (0.001) | 0.018 (0.004) | 0.021** (0.004) | 0.003** (0.003) | 0.008 (0.002) | 0.009 (0.002) | 0.001** (0.001) |
| Strategic Silence (All) | -0.006** (0.002) | 0.000** (0.002) | 0.006** (0.002) | 0.013 (0.002) | 0.016** (0.002) | 0.004** (0.001) | 0.013 (0.001) | 0.015** (0.001) | 0.002** (0.001) | 0.029 (0.003) | 0.032* (0.003) | 0.004** (0.001) | 0.015* (0.001) | 0.015 (0.002) | 0.000* (0.001) |
| Active Speaking (All) | -0.004 (0.013) | -0.006** (0.015) | -0.001** (0.006) | -0.006 (0.006) | 0.001** (0.007) | 0.007** (0.006) | -0.006 (0.004) | -0.003 (0.004) | 0.003 (0.004) | 0.003 (0.007) | -0.011 (0.011) | -0.014* (0.011) | 0.002 (0.003) | -0.007 (0.005) | -0.009* (0.006) |
| Structural Silence (Teaching) | -0.008 (0.002) | -0.007** (0.002) | 0.001** (0.001) | -0.004 (0.001) | -0.004** (0.001) | 0.000** (0.001) | -0.002 (0.001) | -0.001** (0.001) | 0.001** (0.001) | 0.006 (0.002) | 0.014 (0.002) | 0.008** (0.001) | 0.005* (0.001) | 0.009 (0.001) | 0.004 (0.001) |
| Active Speaking (Group Activity) | 0.005 (0.011) | 0.009** (0.012) | 0.004** (0.006) | 0.007 (0.006) | 0.009** (0.006) | 0.002** (0.005) | 0.003 (0.003) | 0.004** (0.004) | 0.001** (0.004) | 0.014 (0.010) | 0.024 (0.009) | 0.010 (0.007) | 0.012* (0.006) | 0.016 (0.005) | 0.004 (0.003) |

Note: H., M., and L. are abbreviations for high-achieving, medium-achieving, and low-achieving respectively. **$p < 0.01$, *$p < 0.05$.

To provide clearer visual representations, the probability density function (PDF) of grouped samples in a specific status was plotted in Figure C1. Fig.C1.a and Fig.C1.b illustrated the distribution of heart rate values across different groups during structural silence during QtDS and strategic silence in response to QtAS, respectively. Notably, the heart rates of low-achieving students differed significantly from those of the other two groups. Fig.C1.c further showed the pronounced differences observed during structural silence in the directed situation. In contrast, as shown in Fig.C1.d, there was no significant variation in the PDF of EDA tonic values among various grades. Fig.C1.e and Fig.C1.f depicted the distribution of RPSD across specific EEG frequency bands for students with varying academic performance during strategic silence following QtDS.

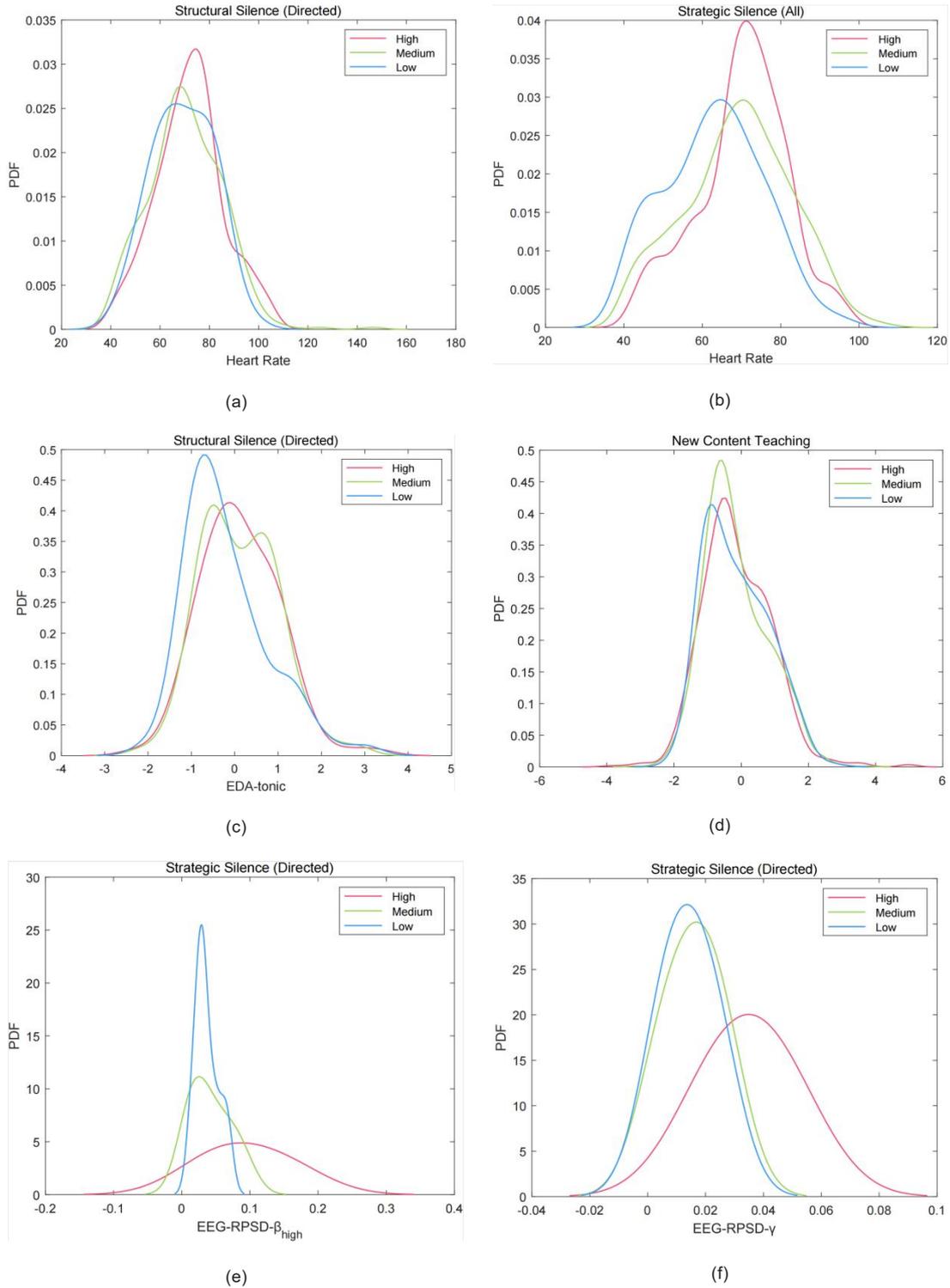

*Figure C1:* PDF of some neurophysiological features for grouped samples in the specific status. **a. and b.** illustrate the distribution of heart rates across different groups during structural silence during QtDS and strategic silence in response to QtAS, respectively. **c.** Pronounced differences in EDA tonic values are displayed during structural silence in QtDS scenarios. **d.** In teaching sessions, there is no significant variation in the PDF of EDA tonic values across different groups. **e. and f.** present the distribution of RPSD across EEG $\beta_{high}$ and $\gamma$ frequency bands for students with varying academic performance during strategic silence following QtDS.